\documentclass[11pt]{article}
\usepackage{amssymb}

\usepackage{amsmath}
\usepackage{graphicx}

\renewcommand{\baselinestretch}{1.1}

\newcommand{\ii}{\'\i}
\begin{document}

\begin{center}
{\large\sc Massive fields temper anomaly-induced inflation:
\\
the clue to graceful exit? }
\vskip 8mm
\textbf{Ilya L. Shapiro}$^{\,a\,,b\,,}$\footnote{%
E-mail: shapiro@dftuz.unizar.es. On leave from Tomsk State
Pedagogical University, Tomsk, Russia.} $\,,\,\,$ \textbf{Joan
Sol\`{a}}$^{\,c\,,d\,,}$\footnote{%
E-mail: sola@ifae.es.} \vskip5mm \vskip 12mm $^{\,a}$
Departamento de F\'{\i}sica Teorica, Universidad de Zaragoza,
\\
50009, Zaragoza, Spain
\vskip 2mm $^{\,b}$ Departamento de
F\'{\i}sica - Instituto de Ciencias Exatas
\\
Universidade Federal de Juiz de Fora, 36036-330, MG, Brazil
 \\
\vskip5mm
{$^{\,c}$\thinspace Departament d'Estructura i
Constituents
de la Mat\`{e}ria, Universitat de Barcelona,\thinspace }\\[0pt]
{Diagonal 647, E-08028, Barcelona, Catalonia, Spain}\\
\vskip2mm $^{\,d}$ {Institut de F\ii sica d'Altes Energies,
Universitat Aut\`onoma de
Barcelona,\\ 08193 Bellaterra, Barcelona, Catalonia, Spain}\\[0pt]
\end{center}

\vspace{0.8 cm}

\begin{center}
{ABSTRACT}
\end{center}

\begin{quotation}
\noindent {\sl A method of calculating the vacuum effective action
for massive quantum fields in curved space-time is outlined. Our
approach is based on the conformal representation of the fields
action and on the integration of the corresponding conformal
anomaly. As a relevant cosmological application, we find that if
taking the masses of the fields into account, then the
anomaly-induced inflation automatically slows down. The only
relevant massive fields for this purpose turn out to be the
fermion fields. So in supersymmetric theories this mechanism can
be specially efficient, for it may naturally provide the graceful
exit from the inflationary to the FLRW phase. Taking the SUSY
breaking into account, the anomaly-induced inflation could be free
of the well-known difficulties with the initial data and also with
the amplitude of the gravitational waves.} \vskip 1mm
\end{quotation}

\vskip 8mm
\newpage

\noindent {\large\bf 1. Introduction} \vskip 1mm

\quad Inflation proved very useful in solving numerous problems of
the theory of the Early Universe. The conventional approach, which
is based on the inflaton field, is extremely helpful for the
inflationary phenomenology \cite{Guth}. At the same time, the
origin of the inflaton remains unclear, and it is quite reasonable
to look for other approaches yielding similar phenomenological
results. Recently, there was an increasing interest for the
cosmological applications of quantum field theory in curved
space-time, specially in connection to vacuum
effects\,\cite{WittenWeinb}.

An exact calculation of the vacuum effective action is
possible only for some special models, usually for massless
conformal invariant matter fields and special restrictions of
the classical background space-time. An important example is the
$d=4$ theory on the conformally flat (or similar FLRW with
$k = \pm 1$) metric, where the Reigert-Fradkin-Tseytlin
effective action \cite{rei} is exact and provides the most
natural theoretical background for inflation
\cite{star,vile,anju}.

The anomaly-induced action can be applied at high energies, where
the masses of the fields are negligible \cite{fhh,vile,anju}. If,
at the high-energy region, there is a supersymmetry (SUSY), one
meets the stable version of the anomaly-induced inflation, which
starts without any fine tuning and leads to sufficient expansion
of the Universe \cite{wave}. In the course of inflation, the
typical energy scale decreases, and one encounters the non-stable
version (Starobinsky model \cite{star}), which provides fast exit
to the FLRW phase \cite{vile}. The transition from stable to
non-stable inflation can be achieved through soft SUSY breaking
and the decoupling of the massive sparticles at low energy
\cite{insusy}. The potential importance of supersymmetry is due to
the following circumstances. The anomaly-induced inflation may be
stable or unstable -- depending on the particle content of the
underlying quantum field theory.
This opens the possibility to interpolate, in a natural way,
between the stable regime at the beginning of inflation and the
unstable regime at the end of inflation. In particular, the
supersymmetric gauge theory may have a particle content
corresponding to the stable inflation. The advantage of stable
inflation is that it starts independent of the initial data for
the conformal factor $a(t)$ of the metric, which can emerge after
the string phase transition. \footnote{We remark that this
mechanism can not explain the homogeneity and isotropy of the
initial state. Perhaps this problem can be solved only in the
framework of the string-inspired inflation.}. If, in the course of
inflation, the typical gravitational energy scale $\mu$ decreases,
and if the sparticles have much bigger masses than the other
particles,
 then they decouple and at the last stage of
inflation the number of active degrees of freedom diminishes. This
is possible because $\mu$ can be sensibly defined from the value
of the Hubble parameter $H$\,\cite{SSCC2}, which indeed lessens
during the tempered expansion caused by massive fields, as will be
shown below. Therefore this mechanism automatically brings
inflation into an unstable phase, with the possibility of an
eventual transition to the FLRW regime.

Thus, supersymmetry and its breaking may provide a natural
qualitative mechanism for the graceful exit from the inflationary
phase, without fine-tuning of the parameters of the theory.
Furthermore, the spectrum of the gravitational waves in this model
\cite{star1,wave} can be in agreement with the existing CMBR data.
Still, in this picture there is an unclear point, namely the use
of the anomaly-induced action for the massive fields is not
completely justified. Also, the transition to the FLRW phase may
not necessary occur after the exit from the exponential inflation.
The behavior of the Universe depends on the initial deviation from
the exponential expansion law \cite{fhh,ander,vile,anju}. It was
established that the universe goes to the FLRW regime if this
deviation leads to an expansion slower than exponential, while it
goes to the uncontrolled ``hyperinflation'' \cite{anju} if it
leads to a faster expansion.  Therefore, it would be very nice to
learn that the masses of the fields really slow down inflation. If
this would be so, the graceful exit to the FLRW phase would not
require any suppositions concerning initial perturbations at the
instant when SUSY breaks down!

In this letter we are going to address both mentioned problems.
We shall develop a simple but reliable Ansatz for the effective
action of massive fields. Our approach to the derivation of the
effective action is based on the Cosmon Model, which was developed
in \cite{PSW} for other purposes-- see also \cite{cosmon2}. The
idea is to construct the conformal invariant formulation
\cite{PSW} of the gauge theory (Standard Model or extensions
thereof, including GUT's),  and then use the well-known methods
to derive the anomaly-induced action. The procedure of
``conformization'' is known for a long time as applied to General
Relativity \cite{deser70} and  Particle Physics \cite{coleman}.
At the classical level, the theory which results from this
procedure is always equivalent to the original theory.
Nevertheless, in the quantum theory the equivalence will be
destroyed by the anomaly, which can be calculated explicitly.
Besides the anomalous terms, there are the conformal invariant
quantum corrections to the classical vacuum action. However, the
complete method of deriving these contributions is not known,
just because the effective action can not be calculated exactly
for the massive $d=4$ theories. The idea of our Ansatz is to
disregard these contributions because they are, indeed, of higher
order with respect to the leading ones we take into account. As
we shall see, our results are in perfect agreement with the
renormalization group. This provides better understanding of the
applicability of our approach.

\vskip 5mm
\noindent
{\large\bf 2. Conformization  and  effective action}
\vskip 1mm

Our first purpose is to construct such a formulation of the
Standard Model (SM) in curved space-time which possesses local
conformal invariance in $d=4$. Actually, the procedure can be
applied to any gauge theory and we are especially aiming at a
realistic supersymmetric gauge theory, providing stability for the
anomaly-induced inflation.

The original action of the theory includes kinetic terms for
spinor and gauge boson fields, as well as interaction terms, all
of them already conformal invariant. As for scalars (e.g. Higgs
bosons) we suppose that their kinetic terms appear in the
combination $\,g^{\mu\nu}\partial_\mu\varphi
\partial_\nu\varphi + 1/6\cdot R\varphi^2\,$ providing the local
conformal invariance. The non-invariant terms are the massive
ones for the scalar and spinor fields:
\begin{eqnarray}
\frac12\int
d^4x\sqrt{-g}\,\,m_H^2\,\varphi^2\,,
\,\,\,\,\,\,\,\,\,\,\,\,\,\,\,\,\,\,\,\,\, \int
d^4x\sqrt{-g}\,\,m\,{\bar\psi}\psi\,.
\label{masses}
\end{eqnarray}
Furthermore, there is an action
for gravity itself, which is also non-invariant
\begin{eqnarray}
S_{EH} =
-\frac{1}{16\pi G}\, \int d^4 x\sqrt{-g}\,R\,. \label{gravity}
\end{eqnarray}

In all mentioned cases the conformal noninvariance
is caused by the presence of dimensional parameters $\,\,m_H^2$,
$\,m$, $\,M^2_P=1/G$. The central idea of the Cosmon Model
\cite{PSW} is to replace these parameters by functions of some new
auxiliary scalar field $\,\chi$. For instance, we replace
\begin{eqnarray}
m_H^2 \to \frac{m_H^2}{M^2}\,\chi^2 \,,\,\,\,\,\,\,\,\, m \to
\frac{m}{M}\,\chi \,,\,\,\,\,\,\,\,\, M_P^2 \to
\frac{M_P^2}{M^2}\,\chi^2\,,
\label{replace}
\end{eqnarray}
where $M$ is
some dimensional parameter, e.g. related to a high scale of
spontaneous breaking of dilatation symmetry \cite{coleman}.
It is supposed that
the new scalar field $\,\chi\,$ takes the values close to $M$,
especially at low energies.
But, there is a great difference between $\,\chi\,$ and  $\,M\,$
with respect to the conformal transformation. The mass does not
transform, while $\,\chi\,$ does. Then, the action of the new
model becomes invariant under the conformal transformation
\begin{eqnarray}
\chi \to \chi\,e^{-\sigma}\,,
\label{auxiliar}
\end{eqnarray}
which is
performed together with the usual transformations for the
other fields
\begin{eqnarray}
g_{\mu\nu}\to
g_{\mu\nu}\,e^{2\sigma} \,,\,\,\, \varphi \to \varphi\,e^{-\sigma}
\,,\,\,\, \psi \to \psi\,e^{-3/2\,\sigma} \,.
\label{conformal}
\end{eqnarray}
It is easy to see that, in the matter field sector, the terms
(\ref{masses}) are replaced, under (\ref{replace}), by Yukawa
and quartic scalar interaction terms. These interactions are between
physical fields (spinors and scalars) and the new auxiliary scalar
$\,\chi$. Thus, in the matter sector our program of
``conformization'' is complete.

However, in the gravity sector (\ref{gravity}) the conformal
symmetry holds only for $\sigma=const$, i.e. only at the level of
global dilatation symmetry, and  this is still not what we need.
Let us make one more step and  require the local conformal
invariance. Then the gravity action must be replaced by the
expression \cite{deser70}
\begin{eqnarray}
S^*_{EH} = -\frac{1}{16\pi
G\,M^2}\,\int d^4 x\sqrt{-g}\, \left[\,R\chi^2 + 6\,(\partial
\chi)^2\,\right]\,,
\label{new gravity}
\end{eqnarray}
where $\,(\partial
\chi)^2=g^{\mu\nu}\partial_\mu\chi
\partial_\nu\chi\,$.
After setting $\,\chi \to M\,$ the expression (\ref{new gravity})
becomes identical to the initial one (\ref{gravity}). This fixing
can be called ``conformal unitary gauge'' in analogy with the
unitary gauge of ordinary gauge theories, and the scale $M$ can be
associated to the vacuum expectation value of the spontaneously
broken dilatation symmetry at high energies
\cite{coleman,PSW,cosmon2}. But, as far as we consider the
space-time dependence of $\chi$ and define its conformal
transformation, the resulting theory exhibits local conformal
invariance under (\ref{auxiliar}) and (\ref{conformal}) with
$\,\sigma =\sigma(x)$. The new conformal symmetry is introduced
simultaneously with the new scalar field $\chi$, which absorbs
the degree of freedom of the conformal factor of the metric. The
new theory satisfies, at the classical level, the conformal
Noether identity
\begin{eqnarray}
\left[\,2\,\frac{g_{\mu\nu}}{\sqrt{-g}} \,
\frac{\delta}{\delta g_{\mu\nu}}
- \frac{\chi}{\sqrt{-g}}\,\frac{\delta}{\delta \chi}
+ d_{\Phi_i}\,\frac{\Phi_i}{\sqrt{-g}}\,
\frac{\delta}{\delta \Phi_i}\,\right]S^c_t= 0\,,
\label{Noether}
\end{eqnarray}
where $\,\Phi_i\,$ stands for the  matter fields of different
spins, $\,\,d_{\Phi_i}\,$ denote their conformal weights and
$\,S^c_t=S^c_t[g_{\mu\nu},\chi,\Phi_i]$ is the total (classical)
action including the modified gravitational term (\ref{new gravity}).

When we quantize the theory, it is important to separate the
quantum fields from the ones which represent a classical
background. In order to maintain the correspondence with the usual
formulation of the SM, we avoid the quantization of the
field $\chi$ which will be considered, along with the metric, as
an external classical background for the quantum matter fields. It
is well known (see, e.g. \cite{book}) that the renormalizability
of the quantum field theory in external fields requires some extra
terms in the classical action of the theory. The list of such
terms includes the nonminimal term of the $\,\int R\varphi^2$-type
in the Higgs sector, and the action of external fields with the
proper dimension and symmetries. The higher derivative part of the
vacuum action has the form
\begin{eqnarray}
S_{vac} = \int d^4 x\sqrt{- g}\, \left\{l_1C^2 + l_2E +
l_3{\nabla^2}R\,\right\}\,,
\label{vacuum}
\end{eqnarray}
where, $l_{1,2,3}$ are
some parameters, $C^2$ is the square of the Weyl tensor and $E$ is
the integrand of the Gauss-Bonnet topological invariant. Now,
since there is an extra field $\,\chi$, the vacuum action should
be supplemented by the $\,\chi$-dependent term. The only possible,
conformal and  diffeomorphism invariant, terms with dimension $4$
are (\ref{new gravity}) and  the$\,\int\chi^4$-term. The last
contributes to the cosmological constant, which we suppose
to cancel and do not consider here in order to keep the discussion
clear and compact. The effect of the cosmological constant will be
reported elsewhere.

The next step is to derive the conformal anomaly in the theory
with two background fields $\,g_{\mu\nu}\,$ and  $\,\chi\,$. Here
we follow the strategy used in a similar situation
\cite{anhesh}. The anomaly results from the renormalization of the
vacuum action \cite{duff} including the terms (\ref{new gravity})
and (\ref{vacuum}). For the sake of generality, let us suppose
that there is also some background gauge field with strength
tensor $\,F_{\mu\nu}$. Then the conformal anomaly has the form
\begin{eqnarray}
<T_\mu^\mu> \,=\, - \,\Big\{
\,wC^2 + bE + c{\nabla^2} R+d F^2
+ f\,[\,R\chi^2 + 6\,(\partial \chi)^2\,] \Big\}\,,
\label{anomaly}
\end{eqnarray}
where $\,w,\,b,\,c\,$ are the $\,\beta$-functions for the
parameters $\,l_1,\,l_2,\,l_3$, and $\,\,f\,\,$ is the
$\beta$-function for the dimensionless parameter $\,\,1/(16\pi
G\,M^2)$ of the action (\ref{new gravity}) which will play an
essential role in our considerations. Finally, $\,d\,$ is the
$\beta$-function for the gauge coupling constant, which is
standard. The values of $\,w,b\,$ and $\,c\,$ depend on the
particle content of the model and are the following (see e.g.
\cite{anju})
\begin{equation}
w=\frac{N_0 + 6N_{1/2} + 12N_1}{120\cdot (4\pi)^2 }\,\,,
\,\,\,\,\,\,\, b= -\,\frac{N_0 + 11N_{1/2} + 62N_1}{360\cdot
(4\pi)^2 }\,\,, \,\,\,\,\,\,\, c=\frac{N_0 + 6N_{1/2} -
18N_1}{180\cdot (4\pi)^2}\,. \label{abc}
 \end{equation}
 Recall that the condition for stable inflation is $c>0$\,\cite{star}. Then one can play with various
 models; e.g. from the previous equation it follows that the particle content  of the
SM ($N_0=4,N_{1/2}=24,N_1=12$) leads to $c<0$ (unstable inflation)
whereas for the Minimal Supersymmetric Standard Model
(MSSM)\,\cite{MSSM} ($N_0=104,N_{1/2}=32,N_1=12$) one has $c>0$
(stable inflation) etc. On the other hand from direct calculation
using the Schwinger-DeWitt method (see e.g. \cite{book} and
references therein) we get
\begin{eqnarray}
f\,=\,\sum_{i}\,\frac{N_i}{3\,(4\pi)^2}\,\frac{m_i^2}{M^2}\,,
\label{f}
\end{eqnarray}
where $N_i$ are the number of Dirac spinors with
masses $m_i$. We note that bosons do not contribute to $f$.

In order to obtain the anomaly-induced effective action, we
put $\,{g}_{\mu\nu} = {\bar g}_{\mu\nu}\cdot e^{2\sigma}\,$ and
$\,\chi = {\bar \chi}\,\cdot e^{-\sigma}\,$, where the metric
$\,{\bar g}_{\mu\nu}\,$ has fixed determinant and the field
$\,{\bar \chi}\,$ does not change under the conformal
transformation. Then, the solution of the equation for the
effective action $\bar \Gamma$ proceeds in the usual way
\cite{rei,book,anhesh}. Disregarding the conformal invariant term
\cite{book} we arrive at the following expression:
$$
{\bar \Gamma} = \int d^4 x\sqrt{-{\bar g}} \,\{w{\bar C}^2 + b
({\bar E}-\frac23 {\bar \nabla}^2 {\bar R})
+ 2 b\,\sigma{\bar \Delta} + d{\bar F}^2 +
$$
\begin{eqnarray}
+ f\,[\,{\bar R}{\bar \chi}^2 + 6\,(\partial {\bar
\chi})^2\,]\,\} \sigma - \frac{3c+2b}{36}\,\int d^4
x\sqrt{-g}\,R^2\,,
\label{quantum}
\end{eqnarray}
which is the quantum
correction to the classical action of vacuum.

Let us compare Eq.(\ref{quantum}) with the quantum correction
from the renormalization group. The expansion of the homogeneous,
isotropic universe means a conformal transformation of the
metric $\,g_{\mu\nu}(t) = a^2(\eta)\,{\bar g}_{\mu\nu},$
where $a(\eta)=\exp\,\sigma(\eta)$ and $\eta$ is the conformal
time. On the other hand, the renormalization group in curved
space-time corresponds to the scale transformation of the metric
$\,g_{\mu\nu} \to g_{\mu\nu}\cdot e^{-2t}\,$ simultaneous with
the inverse transformation of all dimensional quantities
\cite{tmf,book}. For any $\mu$ we have $\mu
\to \mu\cdot e^{t}$. Thus, one can compare the dependence of the
anomaly-induced effective action (\ref{quantum}) on
$\,\,\sigma\,\,$ and  the scale dependence of the
renormalization-group improved classical action. The last is
defined through the solution of the renormalization group equation
for the effective action \cite{tmf,book}
\begin{eqnarray}
\Gamma[e^{-2t}g_{\alpha\beta},{\Phi_i},P,\mu ] =
\Gamma[g_{\alpha\beta},{\Phi_i}(t),P(t),\mu ]\,,
\label{RGEA}
\end{eqnarray}
where $\Phi_i$ is, as before, the set of all fields and  $P$ the
set of all parameters of the theory. In the leading-log
approximation one can take, instead of (\ref{RGEA}), the classical
action and  replace (for the massless conformal theory) $P \to P_0
+ \beta_P\,t$. Now, comparing (\ref{quantum}) with the result of
this procedure, one confirms the complete equivalence of the two
expressions in the terms which do not vanish for $\sigma=const$.
In particular, coefficient $f$ is a factor of the $\beta$-function
for the Newton constant $G$. The important general conclusion is
that the anomaly-induced effective action is a direct
generalization of the renormalization group improved classical
action. On the other hand, the correspondence in the $f$-term
justifies the correctness of our approach and also helps to
learn the limits of its validity.

\vskip 6mm
\noindent
{\large\bf 3. The role of masses in tempering inflation}
\vskip 2mm

In order to understand the role of the particle masses in the
anomaly-induced inflation, let us consider the total action
with quantum corrections
\begin{eqnarray}
S_t = S_{matter} + S_{EH} + S_{vac} + {\bar \Gamma}\,,
\label{totality}
\end{eqnarray}
which does not satisfy the Noether identity (\ref{Noether})
because of the conformal anomaly. We notice that the account of
quantum corrections into the matter sector would be senseless,
because matter and radiation can be treated incoherently as a
fluid. The only important features of the matter action are the
energy density $\rho$, pressure $p$ and their dependence on $a$.
Of course, quantum effects may change these dependencies, but we
can always choose some model for $\,p(a)\,$ and $\,\rho(a)\,$
without going into the details of quantum effects. On the other
hand, as far as we suppose $\,\rho \ll M_P^4\,$ during the
inflation period, the matter-radiation content can not really
affect the expansion of the Universe. Since $\,a(\eta)\,$ grows
very fast during inflation, the energy density greatly decreases
in a very short time and can not play any role. Concerning
pressure, its importance is even smaller, because matter is out
of equilibrium during the inflation.

One of the approximations we made was to disregard higher loop
and non-perturbative effects in the vacuum sector. There is an
attractive possibility to consider the strong interacting regime
using the AdS/CFT correspondence \cite{hhr,KS}, but this goes
beyond the scope of the present letter. Another approximation is
that we take (as the comparison with the renormalization group
shows) only the leading-log corrections. Usually, this is
justified if the process goes at high energy scale. If the
quantum theory has UV asymptotic freedom, the higher
loops effects are suppressed, and our approximation is reliable.
At the low-energy limit, we suppose that the massive fields
decouple and  their contributions are not important. Then Eq.
(\ref{totality}) can be presented in the form
\begin{eqnarray}
&& S_t=\int d^4 x\sqrt{-{\bar g}}\,\Big\{\,
\Big(\,-\frac{M_P^2}{16\pi M^2} + f\sigma\,\Big) \,[\,{\bar R}{\bar
\chi}^2 + 6\,(\partial {\bar \chi})^2\,]\nonumber\\ && -
\Big(\,\frac{1}{4} - d\sigma\,\Big)\,{\bar F}^2\, \Big\} +
S_{matter} +\,high.\, deriv.\, terms\,.
\label{label}
\end{eqnarray}
One can see that the modifications with respect to the case of
free massless fields \cite{anju} are an additional $f$-term and
the contribution to anomaly due to the background gauge fields.

In order to restore the Hilbert-Einstein term and get the
inflationary solution, we fix the conformal unitary gauge
and put $\,\chi = {\bar \chi}\,e^\sigma =M$.
Furthermore, we can choose the conformally flat metric ${\bar
g}_{\mu\nu} = \eta_{\mu\nu}$. Then the gravitational part of the
action (\ref{label}) becomes
$$
S_{grav} = \int d^4x\,\Big\{\,2b\,(\partial^2\sigma)^2
- (3c+2b)\,[(\partial\sigma)^2+ \partial^2\sigma)]^2 -
$$
\begin{eqnarray}
- \,6M_P^2\,e^{2\sigma}\,(\partial\sigma)^2\,
\Big[\,1-\frac{16\pi M^2}{M^2_P}\,f\,\Big] - \Big(\,\frac{1}{4} -
d\sigma\,\Big)\,{\bar F}^2\, \Big\} \,.
\label{flat}
\end{eqnarray}
Computing the equation of motion in terms of the physical time $t$
(where $dt = a(\eta)d\eta$) we find
$$
a^2{\stackrel{....} {a}}
+3a{\stackrel{.} {a}}{\stackrel{...} {a}}
- \left(5 + \frac{4b}{c}\right){\stackrel{.} {a}}^2{\stackrel{..}{a}}
+ a{\stackrel{..} {a}}^2
- \frac{M_{P}^{2}}{8\pi c}\left( a^2 {\stackrel{..} {a}}
+ a{\stackrel{.} {a}}^2\right)+
$$
\begin{eqnarray}
+\frac{2fM^2}{c}\ln a \left( a^2 {\stackrel{..} {a}} +
a{\stackrel{.} {a}}^2\right)+\frac{2fM^2}{c}\,\frac{{\dot a}^2}{a}
- \frac{d{\bar F}^2}{6ca}=0.
\label{for t}
\end{eqnarray}
An exact solution of this fourth order non-linear differential
equation does not look possible, but it can be easily analyzed
within the approximation that $f$ is not too large. Then the new
terms (collected in the second line of Eq. (\ref{for t})) can be
considered as perturbations. Moreover, the last two of them are
irrelevant, because during inflation they decrease exponentially
with respect to the other terms. Thus, in this approximation, the
only one relevant change is the replacement
\begin{eqnarray}
M_P^2 \longrightarrow M_P^2\left[1- \tilde{f}\,\ln a(t)\right] \,,
\label{hawk-limited}
\end{eqnarray}
where for future convenience we have introduced the dimensionless
parameter
\begin{eqnarray}
\tilde{f}\equiv \frac{16\pi f\,M^2}{M_P^2}= \sum_i\,\frac{N_i}{3
\pi}\,\frac{m_i^2}{M_P^2}\,. \label{ftilde}
\end{eqnarray}
Notice that $f$ is given by Eq. (\ref{f}) and so $\tilde{f}$ does
not depend on the scale $M$. Since (\ref{hawk-limited}) is a
slowly varying function, the effect of the masses may be
approximated through the modification of the inflation law
\begin{eqnarray}
a(t) = a_0\,e^{H_1t}\,,\,\,\,\,\,\,\,\,\,\,\,\,\,\,\,H_1=const
\label{starobinsky}
\end{eqnarray}
according to\,\footnote{We remind the reader that the coefficient
$\,b\,$ is negative for any particle content, see eq.(\ref{abc}).}
\begin{eqnarray}
H_1 = \frac{M_P}{\sqrt{- 16\pi b}} \,\,\longrightarrow \,\,
\frac{M_P}{\sqrt{- 16\pi b}}\left[ 1- \tilde{f}\,\ln
a(t)\right]^{1/2}
= H(t)\,,
\label{hawk}
\end{eqnarray}
To substantiate our claim, we have solved Eq. (\ref{for t})
directly using the numerical methods. The plots corresponding to
the numerical solution of the Eq. (\ref{for t}) using
Mathematica\,\cite{Wolfram} are shown in Fig.\,1. Since in the
first period of inflation masses do not play much role and the
stabilization of the exponential inflation proceeds very fast
\cite{anju}, the initial data (in both Eq. (\ref{hawk}) and the
plots of Fig.\,1) were chosen according to the exponential
inflation law:
\begin{eqnarray}
a(0)=1\,,\,\,\,\,\,\,\,\,\,\,\,\,\,\,\, \,\,
{\stackrel{.} {a}}(0) = H_1 \,,\,\,\,\,\,\,\,\,\,\,\,\,\,\,\, \,\,
{\stackrel{..} {a}}(0) = H_1^2\,,\,\,\,\,\,\,\,\,\,\,\,\,\,\,\, \,\,
{\stackrel{...} {a}}(0) = H_1^3\,.
\label{oak}
\end{eqnarray}
\begin{center}
\begin{figure}[tb]
\begin{tabular}{cc}
\mbox{\hspace{1.5cm}} (a) & \mbox{\hspace{0.5cm}} (b) \\
\mbox{\hspace{1.0cm}}\resizebox{!}{4cm}{\includegraphics{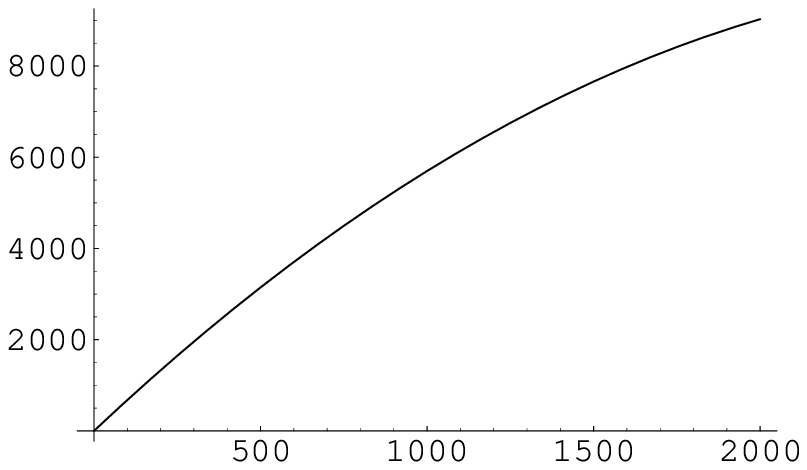}}
& \resizebox{!}{4cm}{\includegraphics{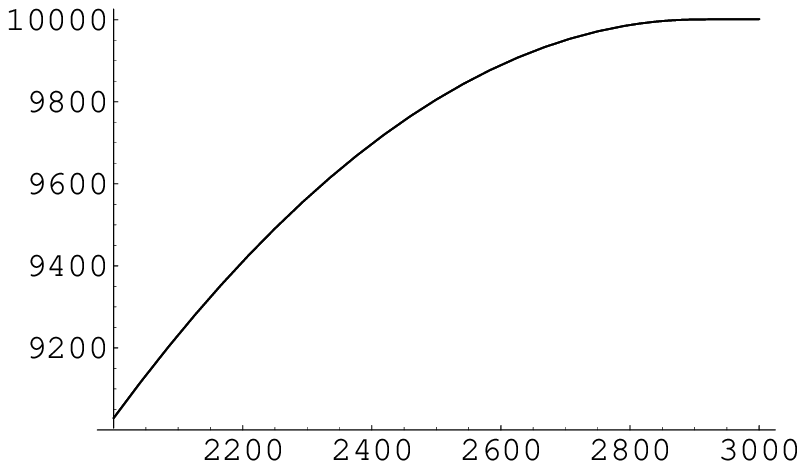}}
\end{tabular}
\begin{quotation}
\noindent \textbf{Figure 1}.\textsl{ \textbf{(a)} Plot of $\ln(a)$
versus the physical time $t$ as a result of the numerical
analysis of Eq.(\ref{for t}); $t$ is given in units of
$16\,\pi/M_P$ and we fixed the parameter (\ref{ftilde}) as
$\tilde{f}=10^{-4}$.
 In this time interval, inflation does not stop, yet; \textbf{(b)} As in
(a), but extending the numerical analysis until reaching an approximate
plateau marking the end of stable inflation.}
 \end{quotation}
 \label{fig1}
\end{figure}
\end{center}
According to the numerical analysis, the total number of $e$-folds
in the ``fast phase'' of inflation (until the Hubble constant
becomes comparable to the SUSY breaking scale) is about $10^4$ for
our particular values of the parameters, and at the last stage the
expansion essentially slows down. The chosen value of the
parameter (\ref{ftilde}) $\tilde{f}=10^{-4}$ in the plot is, as we
warned before, independent of the scale $M$, and it determines
where the process of stable inflation finishes as well as the
number of $e$-folds. Following the above considerations, the
transition from the stable to the unstable inflation can be
associated to a high energy
 scale which we shall
call $M^*$. Let us remark that this typical scale $M^*$ may be
quite different from the scale of SUSY breaking $M_{SUSY}$, in
particular $M^*$ can be some orders of magnitude below $M_{SUSY}$.
The scale $M^*$ is such that there is a sufficient number of
sparticles (scalars and fermions) lighter than $M^*$, so that
$c>0$ even well below $M_{SUSY}$-- see eq.(\ref{abc}). As an
illustration, let us indicate the unique example of the gauge
theory where the spectrum of masses is known: the Standard Model
of particle physics. In the SM the symmetry breaking scale is
given by the vacuum expectation value, $v$, of the Higgs field,
from which one defines the Fermi scale $M_F=G_F^{-1/2} \approx
300\,GeV$. However most of the particles have masses much below
$M_F$ and $v$, and even below $1\,GeV$. One can suppose that a
similar situation takes place in the high energy SUSY GUT. As we
are going to discuss below, the constraint $M^* < 10^{14}\,GeV$
provides better properties of the metric perturbations. It is
important that this does not put rigid limitations on the value of
the scale of supersymmetry breaking which can be some orders of
magnitude greater that $M^*$.

One has to notice that the scale $M_{SUSY}$ and corresponding
$M^*$ are not necessarily linked to a high energy SUSY scale (e.g.
SUSY-GUT, $M_X\sim 10^{16}\,GeV$) but it could just be the SUSY
breaking scale of the MSSM at the $TeV$ scale\,\cite{MSSM}. In the
last case, however, the total number of inflation $e$-folds would
be much greater, but this would not lead to any qualitative change
on the shape of the plots of Fig.\,1 as can be seen from the
analytical structure of eqs.(\ref{for
t})-({\ref{ftilde})\,\footnote{In Fig. 1 we have just illustrated
a situation where the numerical analysis is sufficiently simple,
corresponding to $f$ not too small and so based on a SUSY-GUT
scale. For smaller and smaller $\tilde{f}$ the computer time
becomes exceedingly long. }.

The important qualitative point is that for any value of
$\tilde{f}$ the approximate plateau eventually appears and signals
the end of stable inflation. Also notice from Fig.\,1 that the
initial evolution is close to the exponential inflation
(\ref{starobinsky}), but after that the expansion slows down due
to the quantum effects of massive fermions. The general behaviour
is close to formula (\ref{hawk}). According to the plot in Fig.1
(b), the evolution tends to $H=0$, but before this there must be a
breaking of SUSY and the transition to the unstable phase.
Qualitatively and quantitatively, the plot is in a good
correspondence with formula (\ref{hawk}), especially in the region
$\tilde{f}\,\ln a(t)\ll 1$ where it can be safely used to simplify
the analysis. Remember that the effective action (\ref{quantum})
has been derived in the leading-log approximation, such that the
effect of particles masses has been reduced to the renormalization
of the Newton constant. Indeed, this approximation is valid only
at high energies, when $\,H \gg m_i\,$ for all the fermions. Also,
formula (\ref{hawk}) is based on treating the $\tilde{f}\,\ln
a(t)$-term as a small perturbation.

\vskip 6mm
\noindent
{\large\bf 4. Graceful exit from anomaly-induced inflation}
\vskip 2mm

In order to solve the graceful exit problem in our framework we do
not need to insist that the rate of change of the scale factor,
$\,H(t)$, reduces to zero at some point. Recall that $\,H(t)$ sets
the scale $\mu$ of the renormalization group running for the
gravitational part. If we consider the SUSY breaking and the
corresponding change in the number of active degrees of freedom
\cite{insusy}, then the necessary and sufficient condition for the
applicability of our approach is that $H(t)$ decreases from the
initial value about\,\footnote{Notice that $\,|16\pi b| = {\cal
O}(1)\,$ in the MSSM, and it is much larger than $1$ in any
typical SUSY GUT.} $M_P/\sqrt{-16\pi b}\sim 10^{18}\,GeV$, down to
the lower scale $M^*$. The outcome is that the evolution according
to (\ref{hawk}) lasts until reaching the scale $M^*$, and after
that most of SUSY particles decouple, the inflationary solution
becomes unstable such that the FLRW phase can start. In fact, the
crucial point is the existence of a nonvanishing $f$ as it
eventually tempers stable inflation allowing favorable conditions
for the universe to tilt into the FLRW
phase\,\cite{ander,vile,anju}.

As a result, we arrive at a consistent picture of the graceful
exit from the anomaly-induced inflation to the FLRW stage. It is
easy to see that this conclusion does not change if we choose
another scale for the SUSY breaking. For a lower scale of SUSY
breaking (e.g. $10^{10}\,GeV$ as in the Pati-Salam model, or even
$1\,TeV$ as in the MSSM) there is no need to impose the constraint
on the SUSY spectrum. In this case the area of applicability of
our leading-log approximation is the same as the applicability of
Eq. (\ref{hawk}) and (much more important!) this approximation is
valid until the SUSY breaking scale. The main difference will be
that, for a lower $M_{SUSY}$, the total number of $e$-folds will
increase dramatically, and that the inflation will consume more
time. But, the evolution at the last stage of inflation will be
quite similar as can be seen from Eq.(\ref{hawk}). Another
observation is that, in agreement with (\ref{f}), only spinor
fields contribute to the value of $f$. Therefore, as it was
anticipated in the introduction, taking the masses of the fermion
fields into account we arrive at a tempered form of inflation;
besides, the Universe enters the phase of unstable inflation
\cite{vile,insusy} with such initial conditions that it ends up
with the FLRW behavior. Furthermore, according to (\ref{f}) the
result (\ref{hawk}) is universal, for it does not depend on the
choice of the dilatation symmetry breaking scale $M$. If
interpreted physically, one can put constraints on $M$ using the
macroscopic forces mediated by the field $\sigma$, demanding that
this forces should have the submillimeter range, similarly as in
\cite{SS1}.

Finally, for a really successful exit from the inflation phase we
need to evaluate the dynamics of $H(t)$ during the last 65
$e$-folds of inflation.  The importance of this calculation is
related to the fact that the amplitude of the gravitational waves
\footnote{Let us remind that the spectrum of the gravitational
waves in the exponential phase of inflation is almost flat and
agrees with all observational data \cite{wave}.} is consistent
with the observable range of anisotropy in the CMBR if, during the
last $65$ $e$-folds of the inflation, the Hubble constant $H$ does
not exceed $10^{-5}M_P$. This is because the fluctuations in the
amplitude $h$ of these waves is evaluated using $\delta h/h=H/M_P$
and, on the other hand, is related to the fluctuations in the
temperature of the relic radiation. Thus, it has to satisfy the
relation $\delta h/h=\delta T/T={\cal O}(10^{-5})$. At the lowest
end of the inflation interval this condition corresponds, e.g. in
the SUSY-GUT case to a final scale value $H_f=M^*\lesssim 10^{14}
GeV\approx 10^{-5} M_P$. We expect that after the onset of the
approximate plateau in Fig. 1 (b), where the transition to an
unstable phase occurs, the universe will take a while before
entering the FLRW phase, i.e. the latter will actually initiate at
some point well over the plateau. We have numerically checked that
$H(t)$ decreases very fast on it. For instance, a $15\%$ increase
of the time at the beginning of the plateau amounts $H(t)$ to
diminish two orders of magnitude \,\footnote{This can roughly be
compared (as in the original model\,\cite{Guth}, though of course
in a different sense) to the situation in a supercooled phase
transition in which energy decreases a lot before the transition
really takes place.}. So in general $H(t)$ will decrease further
below $M_{SUSY}$, and the difference between $H_f=M^*$ and
$M_{SUSY}$ at the moment of the transition can be significant, say
one or two orders of magnitude. Hence $M_{SUSY}$ can be
$10^{16}\,GeV$ and this does not create problems with CMBR. Next
we have to derive the value $H_i$ just some number of $e$-folds
$n_e\gtrsim 65$ before the SUSY breaking point $H_f$, where as
usual $n_e$ is defined through $\,a_f/a_i = \exp[n_e].$ We obtain
the following relations:
$$
H_f^2 = H^2_1 + \frac{1}{48\pi^2\,b}\,\sum N_i m^2_i\,\ln a_f\,,
$$
\begin{eqnarray}
H_i^2 = H^2_1 +\frac{1}{48\pi^2\,b}\,\sum N_i m^2_i\,\ln a_i
=H_f^2-\frac{n_e}{48\pi^2\,b}\,\sum N_i m^2_i\,. \label{n_e}
\end{eqnarray}
Notice that $H_i^2>H_f^2$ because $b<0$. However, if we suppose
that $\,H_f=M^*$ and that the sum $\,\sum N_i m^2_i\,$  is of the
order of $\,{M^{*}}^2$, then $H_i$ is of the same order of
magnitude as $H_f$. In other words, the amplitude of the
gravitational waves produced by the anomaly-induced inflation can
be consistent with the magnitude of the CMBR. Remarkably, this
result can be achieved without specifying the details of the gauge
model. It is sufficient to make some  reasonable suppositions
about the mass spectrum of SUSY particles. The numerical analysis
confirms the conclusion derived from the approximate formula
(\ref{n_e}). On the other hand the final conclusion regarding the
consistency with the CMBR observations require an explicit
derivation and analysis of the metric and density perturbations in
the last $65$ $e$-folds of inflation, between $H_i$ and $H_f$.
Such study is beyond the scope of the present considerations, and
it may require an elaborated analysis of the time dynamics of the
Hubble parameter (defining the scale $\mu$ of the gravitational
interactions) in the given fundamental theory. Even at the level
of our relatively simple effective framework, $H(t)$ is obtained
only after numerically solving the non-linear fourth order
differential eq.(\ref{for t}). However, in the gravitational wave
sector, one can make some qualitative observations without
explicit calculations. The corresponding analysis has been
performed, for the case of constant $H$, in \cite{star1} and later
on in \cite{wave} in the effective action framework when all
parameter dependences become explicit. According to this work, the
perturbation spectrum strongly depends on the parameters of the
classical vacuum action and also on the choice of the quantum
vacuum state of the induced theory (which was previously discussed
in \cite{balbi} in relation to the analysis of Hawking radiation
from the black holes). The general conclusion is that the spectrum
is very close to the Harrison-Zeldovich one for the sufficiently
small value of the relevant vacuum parameter $a_1$ (consistent
with the renormalization group) and with the most natural choice
(in comparison to the black hole case) of the quantum vacuum. It
is clear that the same possibilities of changing the perturbation
spectrum exist for the non-constant $H$. Moreover, since in the
phenomenologically important period of inflation the scale factor
changes by more than $65$ $e$-folds while $H(t)$ remains to be the
same order of magnitude, one can suppose that the constant-$H$
terms will dominate in the equations for the metric perturbations
and that the result will not be very different from the one of the
constant $H$ in Ref. \cite{wave}. Therefore, the anomaly-induced
inflation has some predictive power in the description of the
perturbations spectrum. But, the small details of this spectrum
can be changed by adjusting the parameters of the classical vacuum
action and the quantum vacuum. As a result, we may hope to fit
with the present and future experimental data within this model.
In principle, when the amount of such data will become
sufficiently large, one can expect to achieve some additional
information concerning the spectrum of the high-energy theory in
this framework.

\vskip 6mm \noindent {\large\bf 5. Conclusions} \vskip 2mm

In summary, we have considered an effect of particle masses on the
anomaly-induced inflation. The method of calculation was based on
the local Cosmon Model\,\cite{PSW}, i.e. the conformal description
of the massive fields, and on the conventional method of deriving
the anomaly-induced effective action. The output of our approach
agrees with the expressions expected from the renormalization
group. The cosmological application of our result is that,
independent of the details of the particle content of the model,
the (spinor) matter fields slow down inflation. Together with the
supersymmetry breaking effect \cite{insusy}, this provides the
qualitative basis for the graceful exit from the stable
anomaly-induced inflation. Furthermore, there is the possibility
that under certain assumptions concerning the spectrum of the SUSY
GUT, the amplitude of the gravitational waves is consistent with
the CMBR constraints. The precise quantitative description will of
course require to go into the details of a more fundamental theory
(superstring theory or M-theory) underlying this effective
approach.  In the meantime we see that in the anomaly-induced
model there are some indications to a phenomenologically
consistent picture of inflation without introducing an \textit{ad
hoc} inflaton and without fine-tuning the parameters and/or the
initial data.

\vskip 4mm \noindent {\large\bf Acknowledgments.} We are indebted
to V.N. Lukash and A.A. Starobinsky for useful discussions.
Authors are very grateful to Departamento de F\'{\i}sica Teorica,
Univ. de Zaragoza, Departament E.C.M., Univ. de Barcelona, IFAE
(I.L.Sh.), and Departamento de F\'{\i}sica - ICE, Univ. Federal de
Juiz de Fora (J.S.) for warm hospitality and  to CNPq (Brazil) for
support. I.L. Sh. also thanks the finantial support by DGU-MEC
(Spain) and J.S that of CICYT under project No. AEN98-0431, and
the Dep. de Recerca de la Generalitat de Catalunya.
\renewcommand{\baselinestretch}{0.9}

\end{document}